\documentclass[preprintnumbers, floatfix, letterpaper, twocolumn,aps,prd,epsfig,nofootinbib,natbib,longbibliography]{revtex4}

\usepackage{float}
\usepackage{graphicx}
\usepackage{amsmath}
\DeclareMathOperator{\arcsec}{arcsec}

\usepackage{amsfonts}
\usepackage{amssymb,ulem}
\usepackage{color}%
\usepackage[center]{subfigure}
\usepackage{pdfpages}
\usepackage{multirow}

\usepackage{MnSymbol,wasysym}
\usepackage{braket,diagbox}
\usepackage{eurosym}
\usepackage{calrsfs}
\usepackage[usenames,dvipsnames,svgnames]{xcolor}
\allowdisplaybreaks

\newcommand{\RNum}[1]{\uppercase\expandafter{\romannumeral #1\relax}}
\usepackage[colorlinks=true,linkcolor=blue,urlcolor=blue,filecolor=black,citecolor=red,
pdfstartview=FitV,pdftitle={},pdfsubject={},pdfkeywords={},pdfpagemode=None,bookmarksopen=true]{hyperref}

\begin{document}

\title{Trajectories of photons around a rotating black hole with unusual asymptotics}

\author{Yong-Zhuang Li${}^{1}$}
\email{liyongzhuang@just.edu.cn}
\author{Xiao-Mei Kuang${}^{2}$}
\email{xmeikuang@yzu.edu.cn}
\affiliation{ ${}^{1}$ Research center for theoretical physics, School of Science, Jiangsu University of Science and Technology, Zhenjiang, China.}
\affiliation{${}^{2}$Center for Gravitation and Cosmology, College of Physical Science and Technology, Yangzhou University, Yangzhou, 225009, China}


\date{\today}

\begin{abstract}

Most black hole solutions are characterized with asymptotically flat, or asymptotically (anti) de-Sitter behaviors, but some black holes with unusual  asymptotics have also been constructed, which is believed  to provide remarkable insights into our understanding of the nature of gravity. In this paper, focusing on a rotating black hole with unusual asymptotics in Einstein-Maxwell-dilaton (EMD) theory, we innovatively analyze the  photons' trajectories around this black hole background, showing that the unusual asymptotics has significant influences on the photons' trajectories. We expect that our analysis could give more insights in the scenario of black holes' shadow and image.
\end{abstract}

\maketitle


\section{\label{sec:level1}Introduction}

At present, string theory and loop quantum theory appear as two most promising candidates to unify gravity with the other fundamental interactions in nature \cite{Rovelli2008LPG,Oriti2009,Kiritsis2019}. However, due to technical difficulties and conceptual ambiguities, the problem of reconciling quantum theory and general relativity remains open. A complete quantum gravity by one common representation still fascinates the theoretical physics communities. Nevertheless, as the predictions of general relativity, black holes provide natural laboratories for testing different approaches and shedding new light on quantum gravity in the strong field regime. Benefiting from the advances in observational techniques, the more we learn about black holes, the closer we get to a comprehensive theory of quantum gravity.

Over the past few decades, searching the black hole configurations in string and supergravity theory has attracted the attention of many physicists, see  \cite{Myers1986aop,Myers1987npb,Callan1989npb,Garfinkle1991prd,Witten1991prd,Maldacena1996arxiv,Strominger1996plb,Callan1996npb,Maldacena1996npb,Breckenridge1997plb,Emparan2000prl,Mohaupt2000cqg,Maeda2011ptp,Shahbazi2013,Pieri2019} and many other important references. Especially, dilaton black holes have always been a major focus of these investigations, which arise in the low-energy string theory and couple in a nontrivial way to other fields such as gauge fields. Einstein-Maxwell-dilaton (EMD) gravity, where the dilaton is non-minimally coupled to the Maxwell field, then provides a natural theoretical model to investigate the effects of the dilaton on the properties of the black hole spacetime \cite{Garfinkle1991prd,Gibbonsnpb1988,Yoshimura1987plb,Horowitz1992prd,Rakhmanov1994prd,Wiltshire1994prd, Mann1995npb,Sermutlu1995cqg,Soh1998prd, Yazadjiev1999ijmpd, Yazadjiev2001arxiv, Zhang2004prd,Leygnac2004prd, Wang2004prd,Yazadjiev2005cqg, Yazadjiev2005prd,Pakravan2006prd, Aliev2006prd,Ghosh2007prd, Sheykhi2008prd, Sheykhi2008prd2,Soda2009prd, Allahverdizadeh2010grg, Nedkova2016prd, Wu2019jhep, Stetsko2019epjc,Ebrahimkhas2020epjp,Huang2021prd, Ghezelbash2021epjc, Qaemmaqami2022epjp, Fabris2022prd,QiQi:2023nex}.

Especially, Poletti and Wiltshire derived the global properties of static spherically symmetric solutions to the EMD system in the presence of a Liouville-type dilaton potential, showing that with the exception of a pure cosmological constant ‘potential’, no asymptotically flat, asymptotically de Sitter (dS) or asymptotically anti-de Sitter (AdS) solutions exist in these models \cite{Wiltshire1994prd}. Chan et al. then presented a new class of such black hole solutions in $n\geq 4$-dim EMD gravity \cite{Mann1995npb}. Some exact topological black hole solutions in the EMD theory with a Liouville-type dilaton potential were generalized by Cai et al. in Ref. \cite{Soh1998prd}, also see Ref. \cite{Yazadjiev2005cqg}. For the case of rotating black holes, Cl\'{e}ment and Leygnac generated two classes of rotating dyonic black hole solutions with a specialized dilaton coupling parameter $\alpha^{2}=3$ from the non-asymptotically flat magnetostatic or electrostatic black holes in Ref. \cite{Leygnac2004prd}. With one or two Liouville-type dilaton potential(s) in an arbitrary dimension, Cai and Wang looked for and analyzed in details some (static, electrically or magnetically charged) exact solutions of EMD gravity, showing that some of these neither asymptotically flat nor AdS/dS solutions have higher dimensional origins which have well-behaved asymptotics \cite{Wang2004prd}. New 5-dim, non-asymptotically
flat black holes with rotation in one plane were derived by Yazadjiev as a limiting case of new rotating, non-asymptotically flat black ring solutions in a 5-dim EMd theory arising from a six-dimensional Kaluza-Klein theory \cite{Yazadjiev2005prd}. Later, Fabris and Marques exploited the method of gravitational anomaly to compute the temperature of non-asymptotically flat dilatonic black holes \cite{Marques2012epjc}, and the thermodynamics in the canonical and grand canonical ensembles of a class of non-asymptotically flat black holes of the 4D EMD theory with spherical symmetry were analysed in Ref. \cite{Marques2013grg}.

Although such black hole solutions are not as physically well-established as asymptotically flat or AdS/dS solutions, there are still sufficient motivations to conduct detailed investigation and discussion on them as the suggestions in previously mentioned references. On the one hand, it has shown that the methods originally developed for asymptotically flat or AdS/dS black holes, such as the quasi-local energy approach and gravitational anomaly, could be applied to compute the mass, angular momentum or Hawking radiation of non-asymptotically flat rotating black holes \cite{Leygnac2004prd,Marques2012epjc}, see also Ref. \cite{Zhang2002prd} where Zhang et al. checked the mass bound conjecture in de Sitter space with the topological de Sitter solution and its dilatonic deformation, which provides the evidence in favor of the conjecture. On the other hand, it has been widely believed that any consistent theory
of quantum gravity should be holographic \cite{Hooft1993arxiv,Susskind1995jmp}. Indeed, string theory has beautifully demonstrated holography \cite{Witten1998arxiv,Polyakov1998plb,Witten1998arxiv2,Polyakov1998npb,Maldacena1999ijtp}. As the low-energy limits of string theory, holography for EMD theories, including solutions with unusual asymptotic behaviors, have also been set up and investigated in different literature \cite{Kanitscheider2008jhep,Goldstein2010jhep,Chen2010jhep,Gubser2010prd,Charmousis2010jhep,DeWolfe2011prd,Lee2011prd,Taylor2012jhep,Cai2012jhep,Kuang:2012tq,Ghosh2015ass,Park2015prd,Rougemont2017prd,Swingle2018jhep,Goto2019jhep,Mahapatra2019jhep,Jain2020prd,Bayona2020prd,Fu2021prd,Rougemont2023arxiv} and the references therein. 

In this article, we will examine the appearance of the dark silhouette of the slowly rotating EMD black holes with unusual asymptotic behaviors, which still deserves an in-depth investigation. Since the publications of the observations of the Event Horizon Telescope (EHT) on the shadow of the supermassive black holes M87* and SgrA* \cite{Akiyama2019apjl,Akiyama2022apjl}, there have been numerous articles investigating the nature of the shadows of various types of black holes, wormholes, or black branes, which we cannot catalog all here. For EMD theory, Maki and Shiraishi examined the motion of test particles with arbitrary mass, electric charges and dilatonic charges around a static, spherically-symmetric charged dilaton black \cite{Maki1994cqg}; The light paths of normal and phantom EMD black holes were determined analytically by the Weierstrass elliptic functions in Ref. \cite{Azreg2013prd}; Amarilla and Eiroa studied the shadow produced by a spinning Kaluza-Klein black hole in EMD theory \cite{Amarilla2013prd}; In Ref. \cite{Wei2013jcap}, the shadow cast by EMD-Axion black hole and naked singularity were studied, while the geodesic motion of test particles and light in the EMD-Axion spacetime was studied by Flathmann and Grunau \cite{Flathmann2015prd}; The timelike and null geodesics around the static and the rotating (Kerr-Sen dilaton-axion) dilaton black holes were considered and analytically solved in terms of Weierstrass elliptic functions in Ref. \cite{Soroushfar2016prd}; The shadow images of charged wormholes in EMD theory were constructed in Ref. \cite{Amir2019aop}; The light deflection of EM(anti-)D black
holes, using the optical geometry and the Gauss-Bonnet theorem, was studied in Ref. \cite{Jusufi2019aop}, together with the deflection of light by a class of asymptotically flat phantom wormholes; The accretion process in thin disks around static charged and slowly rotating charged dilaton black holes, using the Novikov-Thorne model, was investigated in Ref. \cite{Sepangi2020epjc}; For more investigations, see \cite{Sharma2020ijmpa,Qaemmaqami2022epjc,Sepangi2022prd,Eiroa2023prd}. However, most of these works are focusing on static spherically symmetric, stationary
axisymmetric spacetimes or spacetimes with well-behaved asymptotic, or slowly rotating cases generated from the modified Newman-Jans algorithm. Therefore, there is a need for an extended study of the geodesic properties of spacetimes with unusual asymptotic.

Meanwhile, on the more theoretical side, the photon spheres play a crucial role when one images Einstein rings of black holes in AdS spacetimes or reflects the AdS boundary with the quasinormal modes, which have recently gained significant attentions \cite{Murata2019prl,Murata2020prd,Nambu2020entropy,Tsujimura2021jhep,Zhang2023jhep,Huot2023jhep,Yonezawa2023prr,Li2023arxiv,Yoda2023arxiv,Sun2023arxiv}. Investigations on the null geodesic and photon sphere of slowly rotating EMD black holes with unusual asymptotic could thus uncover a little piece of the new territory to the gauge/gravity correspondence, especially when we have the possibilities to experimentally observe the light path in the framework of analogue gravity \cite{Visser2005lrr,Genov2013nature}.

This paper is organized as follows. In Sec. \ref{sect2}, we start by a briefly introduction to the slowly rotating Einstein-Maxwell-Dilaton black holes. In Sec. \ref{sect3}, by solving the geodesic equation, we study the trajectories of photon on the equatorial plane and on a spherical surface, respectively.  Conclusions and discussions are offered in Sec. \ref{sec:conclusion}.

\section{Einstein-Maxwell-Dilaton theory}\label{sect2}
Following Refs. \cite{Ghosh2007prd,Sheykhi2008prd,Stetsko2019epjc,Huang2021prd}, the action integral for a 4-dimensional gravity with dilaton and linear Maxwell fields can be written in the form
\begin{eqnarray} \label{action}
    S&=&\frac{1}{16\pi}\int_{\mathcal{M}}\text{d}^{4}x\sqrt{-g}\nonumber\\ 
    & & 
    \times\left(R-2\nabla^{\mu}\Phi\nabla_{\mu}\Phi-V\left(\Phi\right)-e^{-2\alpha\Phi}F_{\mu\nu}F^{\mu\nu}\right)\nonumber\\ 
    & &-\frac{1}{8\pi}\int_{\partial \mathcal{M}}\text{d}^{3}x\sqrt{-\gamma}\Theta(\gamma),
\end{eqnarray}
where $R, \,\Phi,\,\alpha$ are the scalar curvature, the dilaton field and the dilaton-electromagnetic coupling parameter, respectively. $V(\Phi)$ denotes the potential which depends on the dilaton field and $F_{\mu\nu}=\partial_{\mu}A_{\nu}-\partial_{\nu}A_{\mu}$ is the electromagnetic filed tensor where $A_{\mu}$ is a component of the electromagnetic potential. The second term in the action (\ref{action}) corresponds to Gibbons-Hawking boundary term. $\partial\mathcal{M}$ denotes the boundary of the manifold $\mathcal{M}$. $\gamma_{\mu\nu}$ is the induced metric on $\partial\mathcal{M}$ and $\Theta$ is the trace of the extrinsic curvature tensor $\Theta_{\mu\nu}$ on $\partial\mathcal{M}$. Varying the action with respect to the gravitational field $g_{\mu\nu}$, the dilaton field $\Phi$ and the electromagnetic potential $A_{\mu}$, one will obtain the field equations in the following forms:
\begin{eqnarray}
    &&R_{\mu\nu}=\frac{g_{\mu\nu}}{2}\left(V(\Phi)-e^{-2\alpha\Phi}F_{\rho\sigma}F^{\rho\sigma}\right)\nonumber\\
    &&\;\;\;\;\;\;\;\;\;\;\;+2\partial_{\mu}\Phi\partial_{\nu}\Phi+2e^{-2\alpha\Phi}F_{\mu\sigma}F^{\sigma}_{\nu};\\
   &&\nabla_{\mu}\nabla^{\mu}\Phi=\frac{1}{4}\frac{\partial V}{\partial \Phi}-\frac{\alpha}{2}e^{-2\alpha\Phi}F_{\rho\sigma}F^{\rho\sigma};\\
   &&\nabla_{\mu}\left(e^{-2\alpha\Phi}F^{\mu\nu}\right)=0.
\end{eqnarray}
The above field equations admits a slowly rotating black hole if some conditions are satisfied:
\begin{eqnarray}
    A_{\varphi}&=&aqh(r)\sin^{2}\theta;\label{elecpotential} \\
    V(\Phi)&=&\Lambda_{0}e^{\tilde{\lambda}_{0}\Phi}+\Lambda e^{\tilde{\lambda} \Phi}, \label{dilatonpotential}
\end{eqnarray}
and $h(r)$ is a function to be defined. The metric is then given by
\begin{eqnarray}\label{metric}
    \text{d}s^{2}&=&-W(r)\text{d}t^2+\frac{\text{d}r^2}{W(r)}-2af(r)\sin^{2}{\theta}\text{d}t\text{d}\phi\nonumber\\
    & &+r^{2}R^{2}(r)\left(\text{d}\theta^2+\sin^{2}{\theta}\text{d}\phi^2\right),
\end{eqnarray}
where 
\begin{eqnarray}
    W(r)&=&-mr^{2\gamma-1}+\frac{(1+\alpha^2)}{(1-\alpha^2)}\left(\frac{r}{b}\right)^{2\gamma}\nonumber \\
    & &-\frac{\Lambda(1+\alpha^{2})^{2}}{2(3-\alpha^2)}\frac{r^{2(1-\gamma)}}{b^{-2\gamma}}+q^2(1+\alpha^2)\frac{r^{2 \gamma-2}}{b^{2\gamma}};\\
    f(r)&=&mb^{2\gamma}r^{2\gamma-1}+\frac{\Lambda}{2}\frac{(1+\alpha^2)^2}{(3-\alpha^2)}\frac{r^{2(1-\gamma)}}{b^{-2\gamma}}\nonumber \\
    & &-q^2(1+\alpha^2)\frac{r^{2\gamma-2}}{b^{2\gamma}}; \\
    \Phi(r)&=&\frac{\alpha}{(1+\alpha^2)}\ln{\left(\frac{b}{r}\right)}; \\
    R(r)&=&e^{\alpha\Phi}; \\
    h(r)&=&r^{-1}.
\end{eqnarray}
The parameters $m,\,q,\, a$ are relating to the black hole's mass, charge and the angular momentum, respectively. Especially, the black hole's physical mass $M$ can be represented by
\begin{equation}
    M=\frac{2b^{2\gamma}\omega_{2}}{16\pi (1+\alpha^2)}m,
\end{equation}
where $\omega_{2}$ is the surface area of a $2$-dimensional unit hypersphere. For other parameters, one could take them in the following form:
\begin{eqnarray}
    \Lambda_0&=&\frac{2\alpha^{2}}{b^2(\alpha^2-1)};\\
    \tilde{\lambda}_0&=&\frac{2}{\alpha};\\
    \tilde{\lambda}&=&2\alpha;\\
    \gamma&=&\frac{\alpha^2}{1+\alpha^2}.
\end{eqnarray}
$b$ is an integration constant that can be set to be unity, while $\Lambda$ is treated as an effective cosmological
constant that appears due to the dilaton potential (\ref{dilatonpotential}). In the limit $\alpha\rightarrow 0$ one will recover the metric functions for a $(n+1)$-dimensional slowly rotating charged Kerr-AdS black hole \cite{Aliev2006prd,Sheykhi2008prd2} while $\alpha=1$ corresponds to the low-energy limit of string theory. In this article we will focus on the weak coupling $\alpha=0.1$.

\section{Null geodesic}\label{sect3}

\subsection{Null geodesics on the equatorial plane}\label{nullequatorial}

As an example, in this section we will investigate the geodesic structure of photons in a slowly rotating EMD spacetime (\ref{metric}). The Lagrangian is written as
\begin{equation}
    \mathcal{L}=\frac{1}{2}\left(g_{tt} \dot{t}^2+g_{rr} \dot{r}^2+2g_{t\phi}\dot{t}\dot{\phi}+g_{\phi\phi}\dot{\phi}^2+g_{\theta\theta}\dot{\theta}^2\right),
\end{equation}
where the dot represents the derivative with respect to an affine parameter $\lambda$. Then two Killing vectors related to the time and azimuthal angle translational invariance of the spacetime provide two conserved quantities, the energy $E$ and the angular momentum $L_{\phi}$ of the photon:
\begin{eqnarray}
     \frac{\partial\mathcal{L}}{\partial \dot{t}}&=&g_{tt}\dot{t}+g_{t\phi}\dot{\phi}\equiv-	 E, \nonumber \\
      \frac{\partial\mathcal{L}}{\partial \dot{\phi}}&=&g_{\varphi\phi}\dot{\phi}+g_{t\phi}\dot{t}\equiv	L_{\phi}
\end{eqnarray}

\begin{figure}[htbp]
\includegraphics[width=0.45\textwidth]{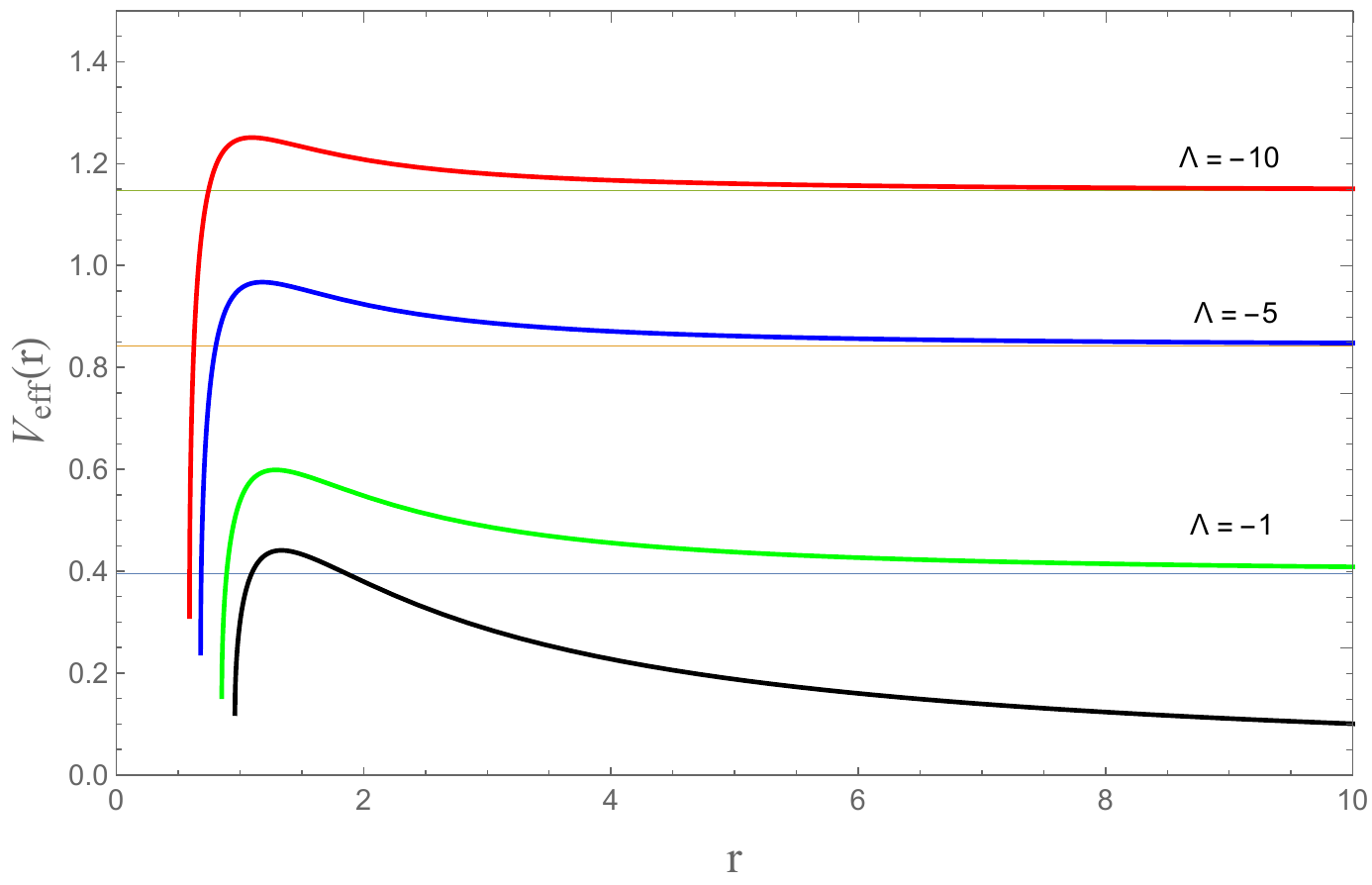}
\caption{\label{fig:diffLambda} The effective potential as a function of $r$ for different $\Lambda$ with $b=1,a=0.1,m=1,q=0.1,\alpha=0.1$. The horizontal lines show $V_{eff}(r\rightarrow\infty)$ for each $\Lambda$. The solid black line stands for $V_{eff}(r)$ with vanishing $\Lambda$.}
\end{figure}

To proceed, one usually needs the third quantity, namely the Carter constant to separate the Euler-Lagrange differential equations for geodesic motion \cite{Carter1968pr}. However, the Jacobi action is inseparable in the present spacetime. To see this, we define the action as 
\begin{equation}
S=-\frac{1}{2}\mu^{2}\lambda-Et+L_{\phi} \phi+\tilde{S}(r,\theta),
\end{equation}
with $$\frac{\partial S}{\partial \lambda}=\frac{1}{2}g^{ij}\frac{\partial S}{\partial x^i}\frac{\partial S}{\partial x^j},$$
one will have
\begin{eqnarray}
   && \left(\frac{\partial\tilde{S}}{\partial \theta}\right)^2+L^{2}_{\phi}\csc{\theta}^{2}+r^{2}R(r)^2W(r)\left(\frac{\partial \tilde{S}}{\partial r}\right)^2\\ \nonumber
   && =\frac{\left(aL_{\phi}f(r)-Er^{2}R(r)^2\right)^2}{a^{2}f(r)^{2}\sin(\theta)^{2}+r^{2}R(r)^{2}W(r)},
\end{eqnarray}
which shows the indivisibility. Indeed, the metric (\ref{metric}) is a Petrov type I spacetime, and violates the so-called Levi-Civita theorem \cite{Civita1904MathAnn,Benenti1980book,Katanaev2023arxiv}.

We now focus on the equatorial plane with $\theta=\pi/2$. In such case, the geodesic equations for photons are given by
\begin{eqnarray}\label{theequatorial}
\dot{r}^{2}&=&\frac{W(r)\left(E^{2}r^{2}R(r)^{2}-2 a E L_{\phi}f(r)-L_{\phi}^{2}W(r)\right)}{a^{2}f(r)^{2}+r^{2}R(r)^{2}W(r)},\nonumber \\
    \dot{t}&=&\frac{E r^2 R(r)^2-a L_{\phi}f(r)}{a^{2}f(r)^{2}+r^{2}R(r)^{2}W(r)},\nonumber \\
    \dot{\phi}&=&\frac{aEf(r)+L_{\phi}W(r)}{a^{2}f(r)^{2}+r^{2}R(r)^{2}W(r)}.
\end{eqnarray}

By eliminating the affine parameter, the relation between $r$ and $\phi$ is written as
\begin{widetext}
\begin{equation}
    \left(\frac{dr}{d\phi}\right)^{2}+\frac{W(r)\left(2 a B f(r)-r^{2}R(r)^{2}+B^{2}W(r) \right)\left(a^{2}f(r)^{2}+r^{2}R(r)^{2}W(r)\right)}{\left(af(r)+BW(r)\right)^{2}}=0,\label{geodesicequation1}
\end{equation}
\end{widetext}
where $B\equiv L_{\phi}/E$ is the impact parameter. By setting a constant radial motion in \eqref{theequatorial}, the effective potential is defined as 
\begin{equation}
    V_{eff}(r)=\frac{W(r)L_{\phi}}{-a f(r)\pm\sqrt{a^{2}f(r)^{2}+r^{2}R(r)^{2}W(r)}},
    \label{effpotential}
\end{equation}
where $``+"$ for prograde motion and $``-"$ for retrograde motion, respectively.

\begin{figure*}[htbp]
  \resizebox{\linewidth}{!}{\begin{tabular}{ccc}
\includegraphics[height=4.cm]{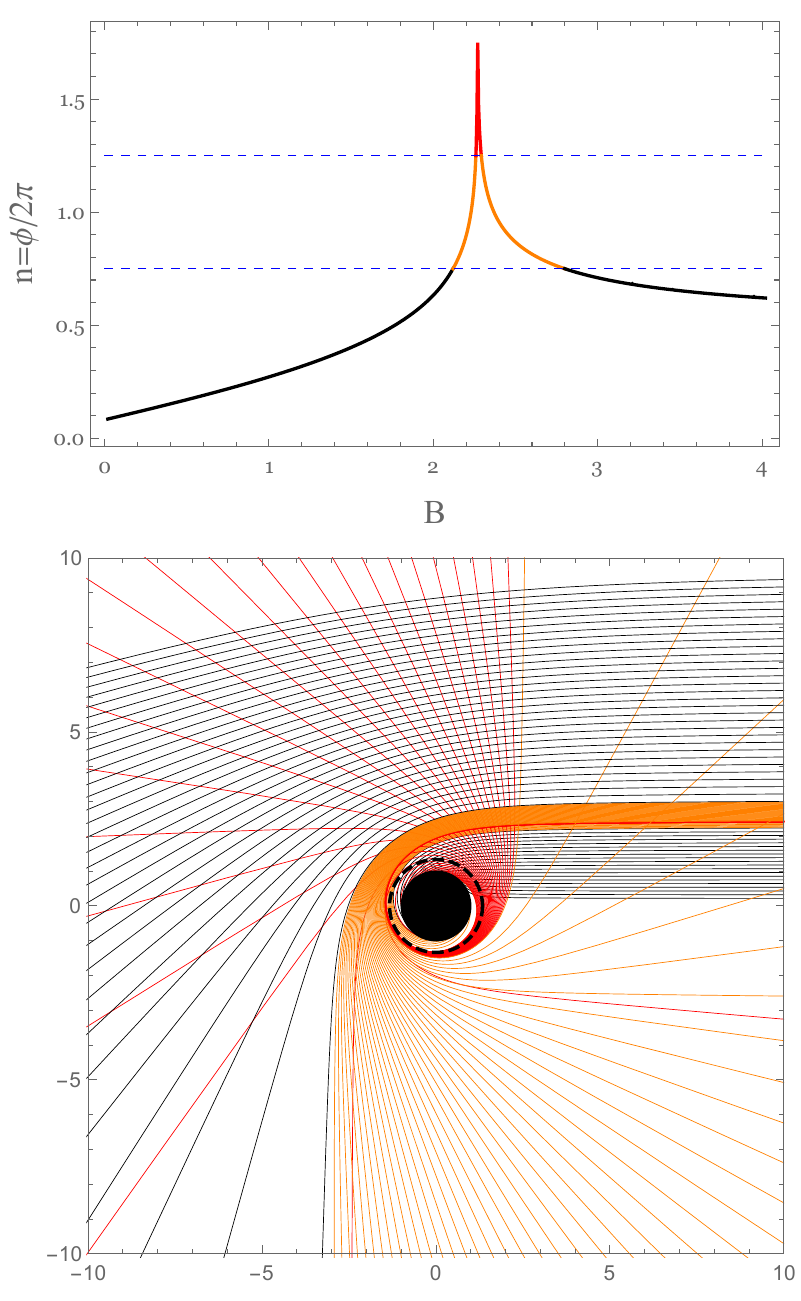}&
\includegraphics[height=4.08cm]{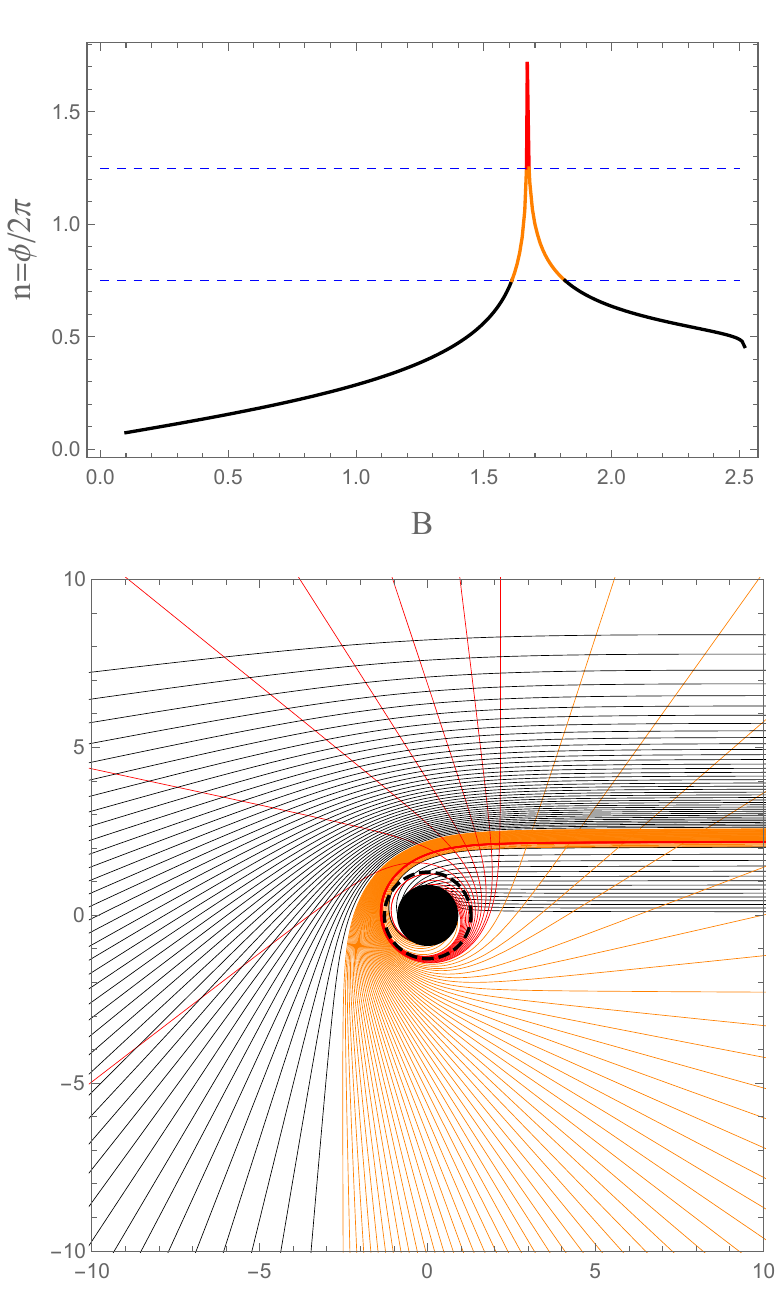}&
\includegraphics[height=4.cm]{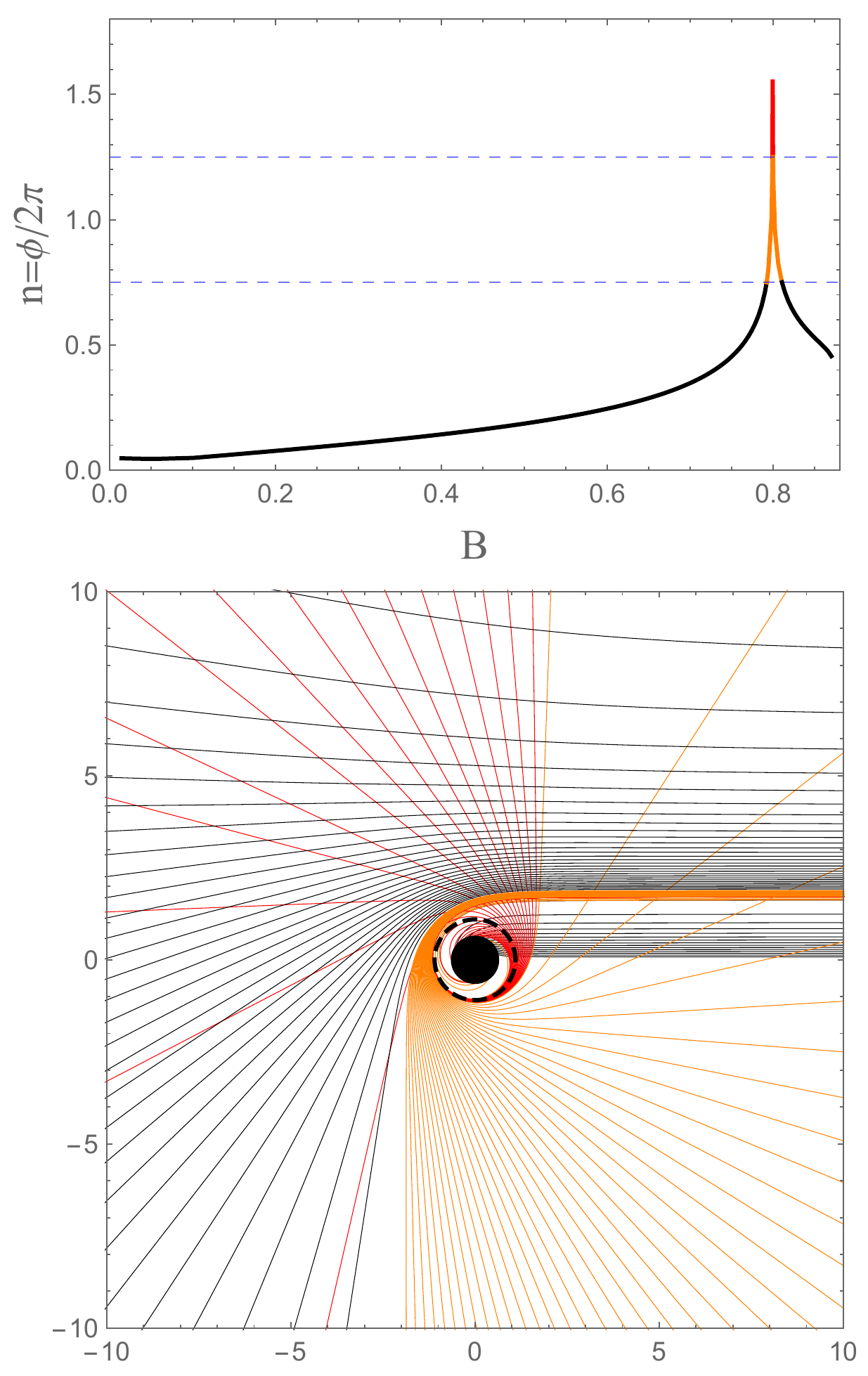}
\end{tabular}}
\caption{\label{fig:nullgeodesic0} The number of photon orbits $n$ (top panel) and trajectories of photons (bottom panel) as a function of the impact parameter $B$ for the slowly rotating EMD black holes with $\Lambda=0$ (left column), $\Lambda=-1$ (middle column) and $\Lambda=-10$ (right column). The black lines, gold lines, and red lines correspond to $(n < 3/4)$, $(3/4 < n < 5/4)$, and $(n > 5/4)$, respectively. The black disk and the dashed curves denote the event horizon and photon sphere.}
\end{figure*}

Fig.~\ref{fig:diffLambda} shows the behavior of the effective potential as a function of $r$ for $\Lambda=-1,\, -5$ and $-10$, respectively. Other parameters have been specifically chosen as $b=1,a=0.1,m=1,q=0.1, \alpha=0.1$, considering the slowly rotating case and weak coupling. The values of these parameters should be restricted to guarantee the existence of black hole horizon, i.e., $W(r)$ should have at least one solution\footnote{Actually for the parameters we have chosen, $W(r)=0$ has two solutions $r_{+}>r_{-}>0$. However, we only consider the photon orbits with radius larger than $r_+$.}. As a comparison, the solid black line describes the behavior of the effective potential when $\Lambda$ is zero but other parameters remain unchanged. Unsurprisingly, the nonzero $\Lambda$ leads to a finite $V_{eff}$ at $r\rightarrow\infty$, while $V_{eff}(r\rightarrow\infty)= 0$ for $\Lambda=0$.

The existence of circular geodesics require $E=V_{eff}(r_{ph})$ and $V_{eff}'(r_{ph})=0$, where $r_{ph}$ is the radius of the photon sphere. With the previous chosen parameters and the numerical method, we plot the number of photon orbits $n$ and trajectories of photons as a function of the impact parameter $B$ for the slowly rotating EMD black holes with parameter $\Lambda=-10, \Lambda=-1$ and $\Lambda=0$ in Fig. \ref{fig:nullgeodesic0}. It should be noted that in our numerical calculation, the light source is located at a very distant position, which can be approximated as being infinity. Following the definitions in Ref. \cite{Gralla:2019xty} the photons are classified into three classes according to $n=\phi/2\pi$: the first class is defined as $n \leq 3/4$, where the deflection angle of the photons is less than $3\pi/2$. The second class with $3/4 < n \leq 5/4$, where light rays deflect with angles $3\pi/2<\phi\leq5\pi/2$. The final class is that the light rays deflect with $n > 5/4$.

One can tell that due to the unusual asymptotics, the trajectories of photons diffuse when the impact parameter $B$ increases. Especially, for the larger absolute value of $\Lambda$, the diffusion is more obvious. Meanwhile, to make sure the square is positive, $B$ is bounded by \eqref{geodesicequation1}. This restriction could be used to understand the repulsive curves shown in the third column of Fig. \ref{fig:nullgeodesic0}. Indeed, for large enough $r$, the critical condition $dr/d\phi=0$ for \eqref{geodesicequation1} could be approximately expressed as
\begin{equation}\label{repulsive1}
    r_{min}^{\zeta}=\frac{\zeta \left[B \Lambda (1+\alpha^2)^{2}(2a-B)+2\alpha^2-6\right]}{4B^{2}(3-\alpha^{2})},
\end{equation}
where $\zeta=-2+4\gamma$. $r_{min}$ is actually the closest point to the black hole for each photon trajectory in the third class. Since $\zeta<0$ this equation will finally give us fixed $B_{max}$ relating to nonzero $\Lambda$, i.e., for nonzero $\Lambda$, the value of $B$ cannot be arbitrarily large. For $\Lambda=0$, \eqref{repulsive1} is simply $r_{min}^{\zeta}=-\zeta/(2B^{2})$. With this in hand, by transforming $(r,\phi)$ into the Cartesian coordinates $(x,y)$ we can always find a ``straight line" $y=-kx+\mathfrak{C}$ where $k>0$ and $\mathfrak{C}=r(\phi=\pi/2)$. This will also lead us a specified $B_{crit}$ for each nonzero $\Lambda$ such that for $B<B_{crit}$ the trajectories bending towards the black hole while for $B_{crit}<B<B_{max}$ the trajectories bending away from the black hole. For the case $\Lambda=0$ there does not exist such a $B_{crit}$, see Appendix \ref{appendixA} for more calculations. This analytical analysis gives the rationality of the diffusion effect and even the repulsive trajectories, but the physics behind these phenomena deserve deeper study.
Moreover, the central shadow region that captures the nearby photon and the widths of $B$ for the second class photons as well as the third class photons are all suppressed by larger absolute value of $\Lambda$.

\begin{figure}[htbp]
\includegraphics[width=0.45\textwidth]{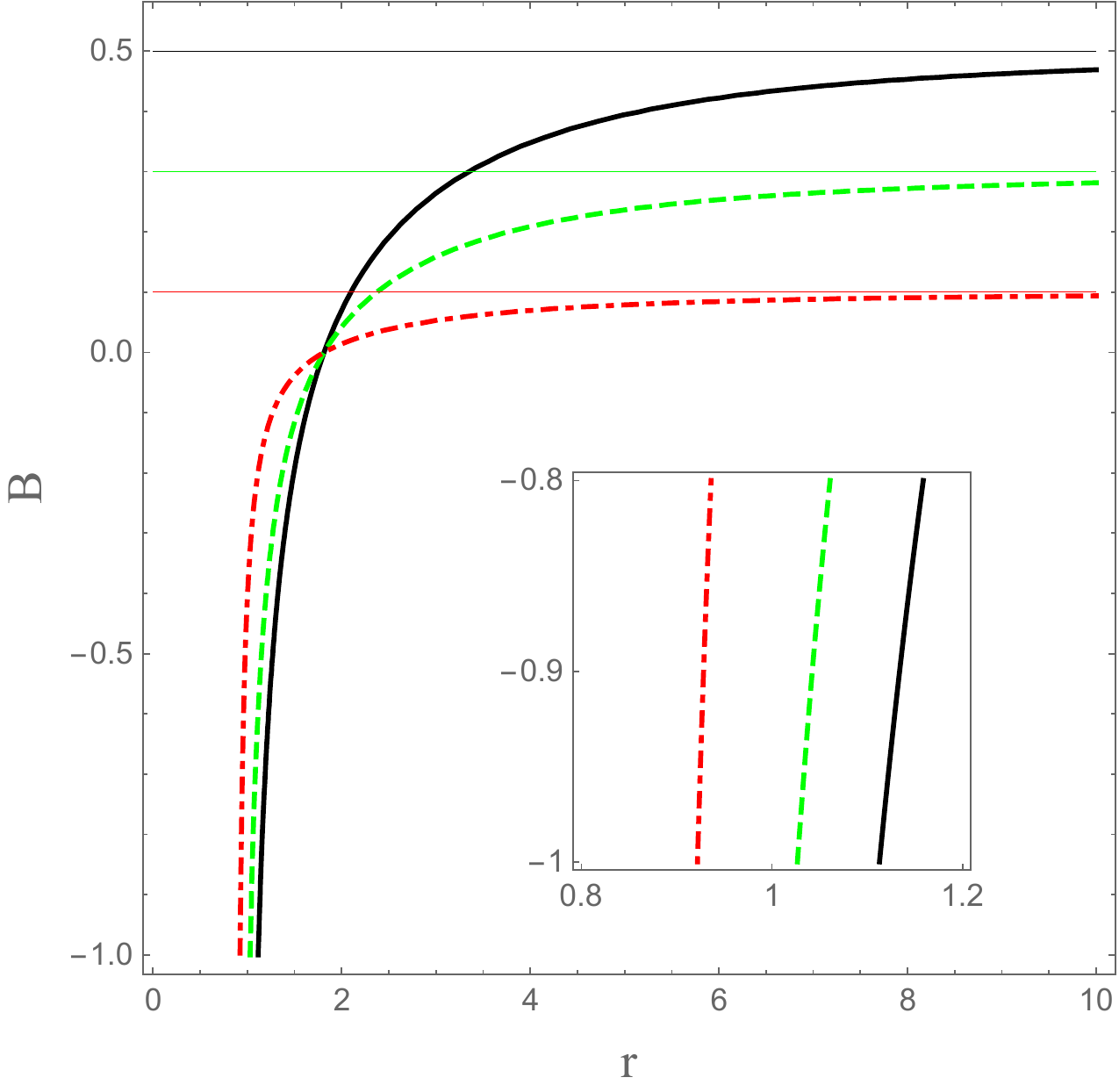}
\caption{\label{fig:aBrrelation} A contour-plot of the condition $af(r)+BW(r)=0$ for $a=0.1$ (red dotdashed), $a=0.3$ (green dashed), $a=0.5$ (solid black), respectively. The horizontal lines show the corresponded $a$ and we set $\Lambda=-1$. The curves are similar for $\Lambda=-10$.}
\end{figure}

\begin{figure}[htbp]
\includegraphics[width=0.45\textwidth]{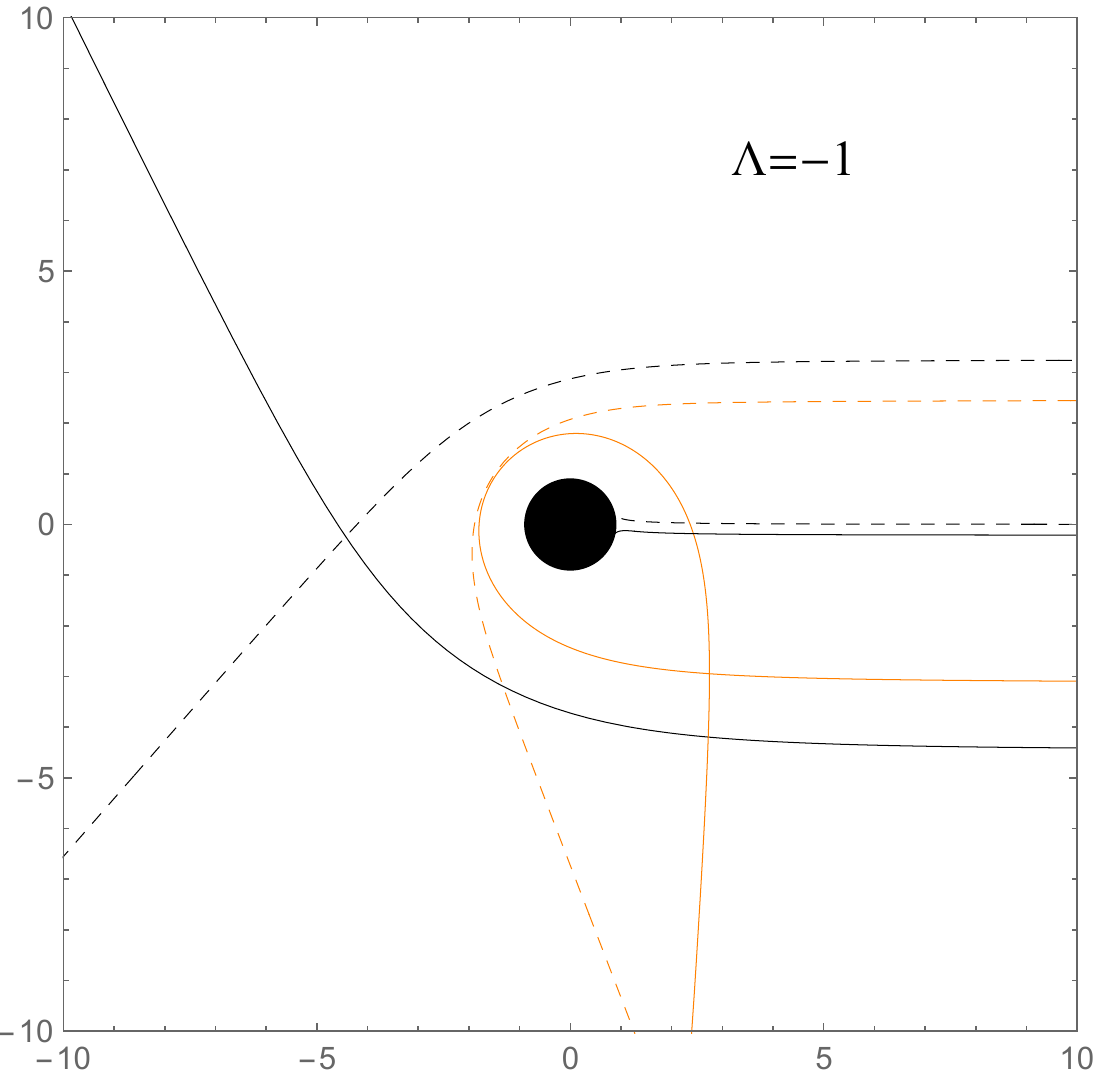}
\caption{\label{fig:comparefandpBi} The photon paths for positive (dashed lines) and negative (solid lines) $B$, respectively. The effective cosmological constant $\Lambda=-1$. $B=\pm 0.1, \pm 1.77, \pm 2$.}
\end{figure}

Another point arises from the relation between $B$ and $a$, see Eq. (\ref{geodesicequation1}) and Fig. \ref{fig:aBrrelation}. For $B>a$, no singularities exist in Eq. (\ref{geodesicequation1}) and the path of photons can fall into the horizon. For $B<a$, the paths have two branches. One will fall into the horizon and the other one will nonphysically ``bounces off".  In particular, for $\Lambda=0$, the situation occurs only for $B<0$, regardless of the values of $a$.
In Fig. \ref{fig:comparefandpBi}, we plot the photon paths for positive and negative impact parameter $B$ with $\Lambda=-1$ as an example. As expected, the paths are different for $B$ with same absolute value but different sign, which is understandable due to the drag effect of the rotating black hole.

\subsection{Null geodesics on the spherical orbits}

Since the black hole is assumed to rotate slowly, performing a perturbative expansion of $a$ is reasonable. Discarding the limitation of $\theta=\pi/2$, and utilizing the Hamilton-Jacobi technologies, the geodesic equations governing the motion of lightlike particles are given by

\begin{eqnarray}\label{orbitsequations}
    \dot{t}&=&\frac{E r^2 R(r)^2-a L_{z}f(r)}{r^{2}R(r)^{2}W(r)},\nonumber \\
    \dot{\phi}&=&\frac{aEf(r)+L_{z}\csc^{2}{\theta}W(r)}{r^{2}R(r)^{2}W(r)},\nonumber \\
    r^{2}R(r)^2\dot{r}^{2}&=&E^{2}r^{2}R(r)^{2}-2aEL_{z}f(r)-W(r)\mathcal{C},\nonumber \\
    &=&\mathcal{R}(r),\nonumber \\
   r^{4}R(r)^4\dot{\theta}^2&=&\left(\mathcal{C}-L_{z}^{2}\csc^2(\theta)\right)=\Theta(\theta), 
\end{eqnarray}
where $\mathcal{C}$ is the Carter constant and we use $L_{z}$ for the angular momentum corresponding to the symmetry axis with different $\theta$. Note for the equatorial plane, $\dot{\theta}=0$, so $\mathcal{C}=L_{z}^{2}$. Substituting the Carter constant in the third equation, one will recover Eq. (\ref{theequatorial}) with small $a$. The spherical orbits will require 
\begin{equation}\label{sphereorbits}
\mathcal{R}(r)=0,\quad \frac{d\mathcal{R}(r)}{dr}=0,
\end{equation}
which generates two equations:

\begin{align}\label{sphericalequation2}
    \frac{\mathcal{C}}{E^{2}}&=\frac{rR(r)\left(rR(r)f'(r)-2f(r)\left(R(r)+rR'(r)\right)\right)}{W(r)f'(r)-f(r)W'(r)},\nonumber \\
    \frac{L_z}{E}&=\frac{rR(r)\left(2W(r)\left(R(r)+rR'(r)\right)-rR(r)W'(r)\right)}{2a \left(W(r)f'(r)-f(r)W'(r)\right)}.
\end{align}
Then one can define the effective inclination angle $i$ as 
\begin{equation}\label{inclinationangle}
    \cos{i}=\frac{L_z}{\sqrt{L^{2}_z+\mathcal{C}}}.
\end{equation}
The radius of polar orbits can be calculated by setting $L_z=0$, i.e., $\cos{i}=0$. While for the equatorial plane, one has $\mathcal{C}=L^{2}_z$ so $\cos{i}=\pm 1/\sqrt{2}$. By eliminating $E$ in equations (\ref{sphericalequation2}), one can obtain the following equation:
\begin{equation}\label{orbitsroots1}
    \mathcal{M}_{1} r^{2}+\mathcal{M}_{2} r+\mathcal{M}_{3}+\mathcal{M}_{4} r^{-1}+\mathcal{M}_{5} r^{-2}+\sum_{n=0}^{2}\mathcal{N}_{n} r^{\zeta-n}=0,
\end{equation}
where $\zeta=-2+4\gamma$, and the detailed expressions of $\mathcal{M}, \mathcal{N}$ are given in Appendix \ref{appendixB}. Thus the radii of generic spherical photon orbits are provided by the positive real roots of this equation.

Finding the analytical solutions to Eq. (\ref{orbitsroots1}) is a difficult task. But one can analytically solve the ordinary differential equations of $\theta$ and $\phi$ for each root of \eqref{orbitsroots1}. With the definition $\mathcal{Z}(\lambda)=\csc{\theta}$, the forth equation in (\ref{orbitsequations}) can be rewritten as 
\begin{equation}\label{newtheta}
\left(\mathcal{Z}'(\lambda)\right)^{2}=-\xi \mathcal{Z}(\lambda)^{2}+(\xi+\eta)\mathcal{Z}(\lambda)^{4}-\eta\mathcal{Z}(\lambda)^{6},
\end{equation}
with
\begin{equation}
    \xi=\frac{vL_{z}^{2}}{(1-v)r^{4}R(r)^4},\quad \eta=\frac{L_{z}^{2}}{r^{4}R(r)^4},
\end{equation}
where $v\equiv\sin^2{i}\in [1/2,1]$. Note that here $r$ is the root of \eqref{orbitsroots1}. The solution to Eq. (\ref{newtheta}) is
\begin{equation}\label{newthetasolution}
    \mathcal{Z}(\lambda)=\frac{\sqrt{\xi}\sqrt{\sec{[\sqrt{\xi}(\lambda-c_{\theta})]}^2}}{\sqrt{\xi+\eta\tan{[\sqrt{\xi}(\lambda-c_{\theta})]}^2}},
\end{equation}
where $c_{\theta}$ is a constant determined by initial conditions. If we set the initial conditions as $\csc\theta_{0}=\mu$, then
\begin{equation}
    c_{\theta}=\pm\frac{1}{\sqrt{\xi}}\arcsec{\left(\frac{\mu}{\sqrt{\xi}}\sqrt{\frac{(\eta-\xi)\xi}{\eta\mu^{2}-\xi}}\right)}.
\end{equation}
Due to the oscillations of $\theta$ about equatorial plan, one can choose either $+$ or $-$. The differential equation of $\phi$ can be recast into 
\begin{equation}
    \phi'(\lambda)=\delta+\zeta \mathcal{Z}(\lambda)^{2},
\end{equation}
with the solution
\begin{equation}
    \phi(\lambda)=c_{\phi}+\int^{\lambda}_{1}\left(\delta+\zeta \mathcal{Z}(y)^{2}\right)dy
\end{equation}
and 
\begin{equation}
    \delta=\frac{aEf(r)}{r^{2}R(r)^{2}W(r)},\quad \zeta=\frac{L_z}{r^{2}R(r)^2}.
\end{equation}
$c_{\phi}$ is also determined by the initial conditions, and by setting $\phi_{0}=0$ one will have
\begin{eqnarray}
    c_{\phi}=\delta&+&\frac{\zeta}{\sqrt{\eta}}\arctan{\left[\frac{\sqrt{\eta}\tan{[\sqrt{\xi}(1+c_{\theta})]}}{\sqrt{\xi}}\right]} \nonumber\\
    &-&\frac{\zeta}{\sqrt{\eta}}\arctan{\left[\frac{\sqrt{\eta}\tan{(\sqrt{\xi} c_{\theta})}}{\sqrt{\xi}}\right]}.
\end{eqnarray}
 Without causing any confusion, we will set the start point at equatorial plan so $\mu=0$, $c_{\theta}=0$ for our numerical calculations.

\begin{figure}[htbp]
\includegraphics[width=0.45\textwidth]{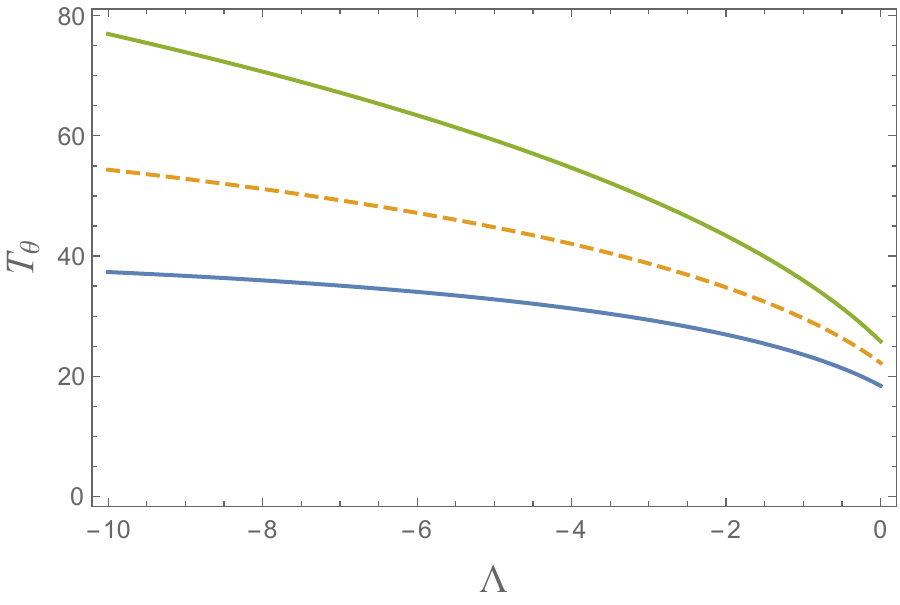}
\caption{\label{fig:vandperiod} The effect of $\Lambda$ on the period $T_{\theta}$ for $v=0.95 (top),\, 0.75 (middle),\, 0.55 (bottom)$, respectively.}
\end{figure}

Eq. (\ref{newthetasolution}) expresses that the oscillation period of $\theta$ around the equatorial plane is $T_{\theta}=2\pi/\sqrt{\xi}$. One can then determine the effect of $\Lambda$ on this period using Eq. (\ref{orbitsroots1}), which has been shown in Fig. \ref{fig:vandperiod}. It is clear that the period $T_{\theta}$ increases as the absolute value of $\Lambda$ increasing. A diffusion-like behavior also happens when we compare the period difference for $v=0.55$ and $0.95$ as $|\Lambda|$ increasing, see Sec. \ref{sec:photonregion} for more discussions. Especially, one can find a specified root of \eqref{orbitsroots1} so $\theta$ and $\phi$ will have same period $2\pi$.

\begin{figure*}
\includegraphics[width=1.0\linewidth]{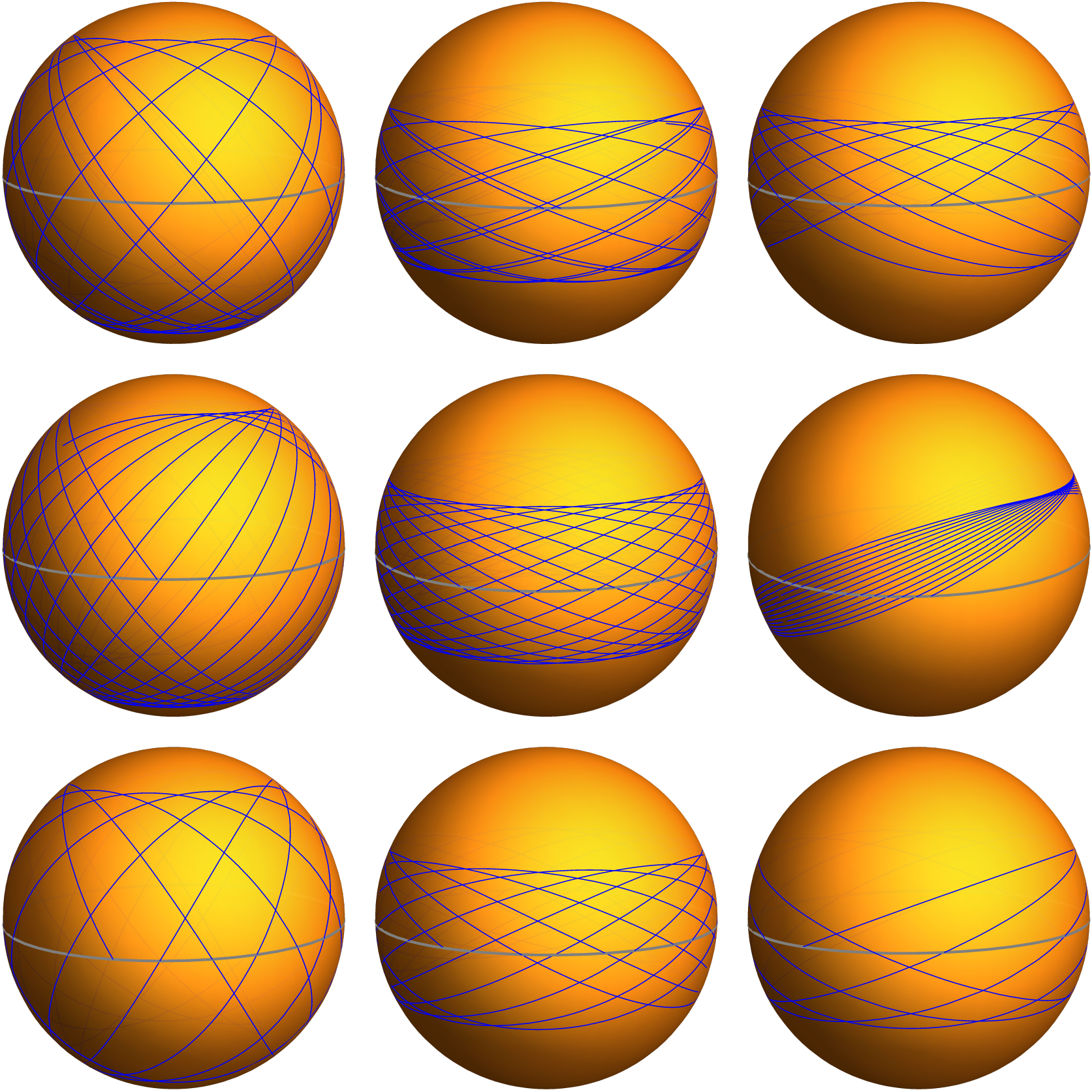}
\centering
\caption{\label{fig:boundedorbits2} The bound null geodesics on the spherical orbits for $\Lambda=0$ (top), $\Lambda=-1$ (middle) and $\Lambda=-10$ (bottom). For each row from left to right $v=0.75,\, 0.55,\, 0.55$, which will fix the radius of the spherical orbits by Eq. (\ref{orbitsroots1}) together with other chosen parameters $\alpha=0.1, a=0.1, m=1,q=0.1, E=0.2$. Especially, the first and second columns correspond to prograde motion, and the third column is retrograde motion, respectively. }
\end{figure*}

{Then we figure out the bound null geodesics on the spherical orbits for $\Lambda=0$ (the top row), $-1$ (the middle row), $-10$ (the bottom row) with $v=0.75$ and $v=0.55$, showing in Fig. \ref{fig:boundedorbits2}. The differences among the plots in each row indicate that the effect of unusual asymptotics is significant. We can also tell that the drag effect reflects in the motions of prograde and retrograde photons by comparing the trajectories between the second and third columns.}

\subsection{Photon regions}
\label{sec:photonregion}

The region accommodating the photon orbits that stay on a sphere constitute the so-called photon region. Besides the conditions (\ref{sphereorbits}), the photon region also needs the condition $\Theta(\theta)\geq 0$, which leads to
\begin{widetext}
\begin{equation}
    rR(r)f'(r)-2f(r)\frac{d(rR(r))}{dr}\leq \csc{\theta}^{2}\frac{\left(2rW(r)R'(r)+2W(r)R(r)-rR(r)W'(r)\right)^{2}}{4a^{2}\left(W(r)f'(r)-f(r)W'(r)\right)},
\end{equation}
\end{widetext}
where the prime denotes a derivative to $r$. The stable or unstable spherical null geodesics in this region are determined by the sign of $\mathcal{R}''(r)$. The condition $\mathcal{R}''(r)>0$ indicates that the spherical null geodesics are unstable, while $\mathcal{R}''(r)<0$ indicates the stable spherical null geodesics. We consider only the photon region outside the black hole horizon. 

\begin{figure}[htbp]
\includegraphics[width=0.45\textwidth]{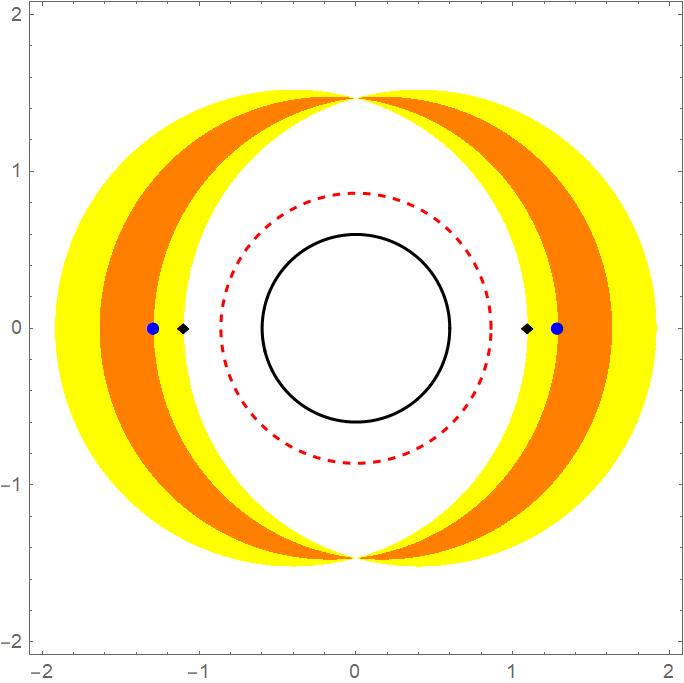}
\caption{\label{fig:photonregion} The photon region outside the horizon for $\Lambda=-1$ (orange) and $\Lambda=-10$ (yellow). The red dashed line and black solid line denote the horizons for $\Lambda=-1$ and $\Lambda=-10$, respectively. The blue circles and dark diamonds signify the photon spheres on the equatorial plan with $\Lambda=-1$ and $\Lambda=-10$, respectively.}
\end{figure}

Fig. \ref{fig:photonregion} shows the photon regions for $\Lambda=-1$ and $-10$. With smaller horizon radius, the photon region becomes larger. In other words, a negative $\Lambda$ with larger absolute value will cause a larger photon region. This feature also explains the behavior in Fig. \ref{fig:vandperiod}. In fact, the spherical orbits are more closer to black hole horizon for larger $\Lambda$, resulting in a shorter period for zenith angle $\theta$.

\section{Closing remarks}\label{sec:conclusion}

{In this article, we analytically and numerically studied the null geodesics around a slowly rotating black hole with unusual asymptotics in Einstein-Maxwell-Dilaton gravity. In order to disclose the effects of unusual asymptotic behaviors on the trajectories of photons, we, as a first attempt, trace the photons' motions on the equatorial plane and on the spherical surface, respectively.

Comparing to asymptotically flat spacetime, we found some interesting features for the trajectories  photons on the equatorial plane of the rotating black hole with unusual asymptotics denoted by a effective cosmological constant $\Lambda$. Firstly, the effective potential of the photons asymptotically approaches a finite constant depending on $\Lambda$, instead of zero in the asymptotically flat case. Secondly, the  unusual asymptotics make the trajectories of photons diffuse and this diffusion is more significant for the larger absolute value of $\Lambda$, which corresponds to that  the central shadow region  and the widths of $B$ for the second class photons as well as the third class photons all become smaller as the absolute value of $\Lambda$ increases. In addition, we observed the repulsive trajectories of the photon in the case with nonzero $\Lambda$, of which we give the analytical understanding from the equation of motion. We believe that the deeper physics behind these novel features deserve further investigation, and the work in this direction is in progress. 

Then, we extended the study into spherical orbits by performing a perturbative expansion on $a$ so we can let $\theta$ be free. We found that  it is tough to analytically solve the compete null geodesic equations, but we can fix the trajectories in a spherical surface and analytically solve the equations for $\theta$ and $\phi$. So we managed to plot the bound trajectories of photons
on the spherical surface, where the prints of unusual asymptotics are explicit. Finally, we figured out the photon regions of this rotating black hole with unusual asymptotics. 
Though the black hole with a negative $\Lambda$  with larger absolute value  have smaller event horizon, it has a wider photon region due to the diffusion of light rays.

The study of null geodesic and the trajectories of photons around a central black hole is the first step to explore the black hole shadow and images. Based on our results, it is interesting to extend into the shadows related physics for the black hole with unusual asymptotics in this scenario and beyond. We also expect that this direction somehow has potential connection with EHT observations.}

\begin{acknowledgments}
We appreciate Zi-Liang Wang for helpful discussions. This work is partly supported by Natural Science Foundation of China under Grants No.12375054 and Natural Science Foundation of Jiangsu Province under Grant No.BK20211601.
\end{acknowledgments}

\appendix
\section{Calculations for $B_{crit}$}\label{appendixA}
\setcounter{equation}{0}
\renewcommand\theequation{A.\arabic{equation}} 

We can use one coordinate transformation, two equations and three points to determine $B_{crit}$. 

\begin{itemize}
    \item To proceed, we first transform the polar coordinate $(r, \phi)$ into the Cartesian coordinate $(x, y)$: $x=r \cos{\phi}$, $y=r \sin{\phi}$. Therefore, if there really exist different types of null geodesics with different deflections, there must be a critical photon trajectory that can be represented in the Cartesian coordinates as $y=-kx+\mathfrak{C}$. Then one will have the relation 
    \begin{equation}\label{appAslope}
        \frac{dy}{dx}=\frac{x+r'(\phi) \sin{\phi}}{-y+r'(\phi) \cos{\phi}}=-k.
    \end{equation}
    \item Suppose the observer is located at $\mathcal{S}_{o}$ on the $x$-axis, and the impact parameter of the critical trajectory is $B$, then we have the first relation
    \begin{equation}\label{appABSC}
        B=-k\mathcal{S}_{o}+\mathfrak{C},
    \end{equation}
    where $\mathcal{S}_{o}$ is taken to be finite but large enough in our calculations. 
    \item Consider $\phi=\pi/2$, which corresponds to $x=0$. Then $y=\mathfrak{C}=r(\pi/2)$. Using \eqref{appAslope} and \eqref{geodesicequation1} one will get the second relation:
    \begin{widetext}
    \begin{equation}\label{appAxis0}
        -\frac{W(\mathfrak{C})\left(2 a B f(\mathfrak{C})-\mathfrak{C}^{2}R(\mathfrak{C})^{2}+B^{2}W(\mathfrak{C}) \right)\left(a^{2}f(\mathfrak{C})^{2}+\mathfrak{C}^{2}R(\mathfrak{C})^{2}W(\mathfrak{C})\right)}{\left(af(\mathfrak{C})+BW(\mathfrak{C})\right)^{2}}=k^{2}\mathfrak{C}^{2}.
    \end{equation}
    \end{widetext}
    The left-hand side of this relation should be positive to make it true. So the polynomial in the first bracket needs to be negative. As $B$ is limited for nonzero $\Lambda$ so $\mathfrak{C}$ is also finite.
    \item The condition $r'(\phi)=0$ actually indicates the closet point $r_{min}$ to the black hole, leading us to the third relation:
    \begin{equation}\label{appArmin}
        2 a B f(r_{min})-r_{min}^{2}R(r_{min})^{2}+B^{2}W(r_{min})=0.
    \end{equation}
    Then using \eqref{appAslope} we can find the Cartesian coordinates for this point, i.e., 
    \begin{eqnarray}
        x_{min}&=&r_{min}\cos{\phi_{min}}=\frac{k\mathfrak{C}}{1+k^2},\nonumber \\
        y_{min}&=&r_{min}\sin{\phi_{min}}=\frac{\mathfrak{C}}{1+k^2},
    \end{eqnarray}
    which represents the fourth relation
    \begin{equation}\label{appArminC}
        r_{min}=\frac{\mathfrak{C}}{\sqrt{1+k^2}}.
    \end{equation}
\end{itemize}
Now we have four relations \eqref{appABSC}, \eqref{appAxis0}, \eqref{appArmin} and \eqref{appArminC} for four unknown parameters $(B,k,\mathfrak{C}, r_{min})$.

However, for $\Lambda=0$, as been discussed in Sec. \ref{nullequatorial}, the relation \eqref{appArmin} will not provide an upper bound on $B$ for large enough $r$. In other words, $B$ increases as $r$ increases. This feature violates the relation \eqref{appAxis0} as $B$ could be arbitrarily large, which indicates that the assumed straight line $y=-kx+\mathfrak{C}$ does not exist.

\section{Coefficients}\label{appendixB}
\setcounter{equation}{0}
\renewcommand\theequation{B.\arabic{equation}} 

With the definition $v=\sin^2{i}$, the coefficients of Eq.(\ref{orbitsroots1}) are
\begin{align}
    \mathcal{M}_{1}&=-4v, \\
    \mathcal{M}_{2}&=4 a^{2}m(v-1)(1+\alpha^2)\Lambda-\frac{4mv(\alpha^{2}-3)}{\alpha^{2}+1}, \\
    \mathcal{M}_{3}&=16q^{2}\left(\frac{a^{2}(v-1)(1+\alpha^{2})^2\Lambda}{\alpha^{2}-3}-v\right)-\frac{m^{2}v(\alpha^{2}-3)^2}{(1+\alpha^2)^2}, \\
    \mathcal{M}_{4}&=-\frac{8mq^{2}v(\alpha^{2}-3)}{a+\alpha^2}, \\
    \mathcal{M}_{5}&=-16vq^{4}, \\
    \mathcal{N}_{0}&=\frac{4a^{2}m^{2}(v-1)(\alpha^{2}-3)}{1-\alpha^{2}}, \\
    \mathcal{N}_{1}&=\frac{8a^{2}mq^{2}(v-1)(5-\alpha^{2})(1+\alpha^2)}{1-\alpha^2}, \\
    \mathcal{N}_{2}&=\frac{32a^{2}q^{4}(1-v)(1+\alpha^2)^2}{1-\alpha^2}. 
\end{align}


\end{document}